\documentclass[journal]{IEEEtran}
\ifCLASSINFOpdf
\else
   \usepackage[dvips]{graphicx}
\fi
\usepackage{url}
\IEEEoverridecommandlockouts
\usepackage{cite}
\usepackage{amsmath,amssymb,amsfonts}
\usepackage{algorithm}
\usepackage{algpseudocode}
\usepackage{graphicx}
\usepackage{booktabs}
\usepackage{textcomp}
\usepackage{xcolor}
\usepackage{siunitx}
\usepackage{multirow}
\usepackage{soul}
\def\BibTeX{{\rm B\kern-.05em{\sc i\kern-.025em b}\kern-.08em
T\kern-.1667em\lower.7ex\hbox{E}\kern-.125emX}}
\begin{document}
\title{EGGCodec: A Robust Neural Encodec Framework for EGG Reconstruction and F0 Extraction
\thanks{$^{\dag}$ These authors contributed equally. $^{\star}$ Corresponding author. This work is supported in part by the National Social Science Foundation of China (23AYY012) and USTC (YD2110002303). Rui Feng, Yu-Ang Chen, Yu Hu, Jun Du, and Jia-Hong Yuan are with the National Engineering Research Center of Speech and Language Information Processing, University of Science and Technology of China (USTC), Hefei, P. R. China. Email: \{fengruimse, yuangchen21\}@mail.ustc.edu.cn, yuhu@iflytek.com, \{jundu, jiahongyuan\}@ustc.edu.cn.
}}
\author{Rui Feng$^{\dag}$, \textit{Graduate Student Member, IEEE}, Yuang Chen$^{\dag}$, \textit{Graduate Student Member, IEEE}, Yu Hu,\\ Jun Du, \textit{Senior Member, IEEE}, Jiahong Yuan$^{\star}$, \textit{Member, IEEE}}

\maketitle
\begin{abstract}
This letter introduces EGGCodec, a robust neural Encodec framework engineered for electroglottography (EGG) signal reconstruction and F0 extraction. We propose a multi-scale frequency-domain loss function to capture the nuanced relationship between original and reconstructed EGG signals, complemented by a time-domain correlation loss to improve generalization and accuracy. Unlike conventional Encodec models that extract F0 directly from features, EGGCodec leverages reconstructed EGG signals, which more closely correspond to F0. By removing the conventional GAN discriminator, we streamline EGGCodec's training process without compromising efficiency, incurring only negligible performance degradation. Trained on a widely used EGG-inclusive dataset, extensive evaluations demonstrate that EGGCodec outperforms state-of-the-art F0 extraction schemes, reducing mean absolute error (MAE) from 14.14 Hz to 13.69 Hz, and improving voicing decision error (VDE) by 38.2\%. Moreover, extensive ablation experiments validate the contribution of each component of EGGCodec.
\end{abstract}

\begin{IEEEkeywords}
F0 extraction, EGG reconstruction, speech inverse filtering, speech signal processing.
\end{IEEEkeywords}

\section{Introduction}
\par Fundamental frequency (F0) extraction is a foundational task in speech signal processing since it reflects the rate of vocal fold vibration and carries essential information pertinent to prosody and speaker characteristics \cite{10800188}. The accurate extraction of F0 is thus essential for various practical applications, including speech recognition, speech synthesis, speaker identification, prosody analysis, and music research \cite{yang2023mixed,10095120,9628174,10508448,berti2023fundamental}. 

\par Currently, there are numerous studies have been conducted on F0 extraction from speech signals. To improve F0 extraction accuracy, the authors in \cite{10508448} proposed an auditory gain harmonic detection method exploiting selective Gammachirp filters to highlight harmonics and reduce the masking effects of acoustic noise. In \cite{DBLP}, the authors focused on melody extraction, which addresses harmonic resolvability that is a key challenge in F0 extraction. The authors in \cite{gaznepoglu2023deep} presented a deep learning approach for synthesizing F0 trajectories for speaker anonymization, addressing noise sensitivity and computational complexity. In \cite{10800188}, we proposed a novel F0 extraction approach named Wav2F0, which combines the Wav2vec 2.0 model with fully connected layers and LSTM and leverages the pre-trained representations learned by Wav2vec 2.0 \cite{10800253}. The authors in \cite{kim2018crepe} proposed a data-driven F0 extraction algorithm named CREPE, which is based on deep convolutional neural networks (CNN) and directly operates on time-domain waveforms. Although the multitude of techniques that have emerged for F0 extraction, the intricate vibration mechanisms of the vocal folds, coupled with the variability of recording conditions, render the task of accurately extracting F0 from waveforms a formidable challenge. Unlike microphone-captured speech signals, EGG signals provide higher accuracy and stability, as they more precisely reflect the periodic nature of vocal fold vibrations, making them more suitable for F0 extraction \cite{herbst2020electroglottography}. EGG is currently the most commonly used visual inspection method for indirectly observing vocal cord vibration, which characterizes vocal cord vibration by measuring the change in resistance between two electrodes placed on the skin covering the thyroid cartilage \cite{nguyen2023exploring, henrich2004use}.

\par In this context, META Corporation in \cite{defossez2022high} developed an innovative neural network-based speech codec called \textbf{Encodec}, which effectively compresses and reconstructs speech signals. Encodec features a symmetric decoder and a convolutional encoder that compress and restore speech signal layers through CNN. Furthermore, Encodec employs generative adversarial networks (GAN) that have been widely used to enhance the perceptual quality of synthesized audio by employing a discriminator to distinguish between real and generated signals \cite{kumar2019melgan}. Building on this approach, Encodec utilizes a GAN-based framework to improve the quality of reconstructed audio, ensuring more natural and realistic speech output. Due to the strong capability of Encodec in maintaining high audio quality and fidelity, Encodec is widely used in studies related to F0 and EGG signal processing. For example, the author in \cite{10974362} proposed a high-fidelity speech codec, which uses both speech and electroglottographic (EGG) signals for speech compression. The authors in \cite{10447523} introduced a fundamental neural speech codec toolkit called FunCodec, which delivers comparable speech quality with reduced computational and parameter complexity. For time-domain models, the authors in \cite{10447523} employed the same SEANet architecture as Encodec. The authors in \cite{10446672} proposed EnCLAP for automatic audio captioning, which integrates two acoustic representation models, i.e., EnCodex and CLAP, with a pre-trained language model BART. In addition, the authors in \cite{10577150} presented a multi-functional speech generation model named SpeechX, which uses Encodec as its neural encoder to handle both clean and noisy signals for zero-shot TTS and various speech transformation tasks. Encodec is primarily adopted for speech reconstruction, and the advantages of its input-output alignment and stable reconstruction quality have enabled it to demonstrate good EGG signal fitting in preliminary experiments. However, due to the complexity of Encodec's structure, limitations in the loss function, and the inherent instability of the GAN discriminator during practical applications, directly using the Encodec framework to reconstruct EGG signals is challenging.

\par This letter designs a robust neural Encodec framework called EGGCodec for precise EGG reconstruction and F0 extraction. EGGCodec leverages the superior ability of EGG signals to characterize vocal fold vibrations and locate them as the reconstruction target for F0 extraction. The structure and training processing of EGGCodec have been greatly simplified by removing the GAN discriminators. To accurately measure the similarity between the reconstructed and target signals, a multi-scale frequency-domain loss function is proposed to capture the nuanced relationship between original and reconstructed EGG signals, complemented by a time-domain correlation loss to improve generalization and accuracy. Extensive evaluations demonstrate that EGGCodec outperforms state-of-the-art schemes across multiple performance metrics.

\vspace{-0.4em}

\begin{figure}[h]
\centering
\includegraphics[scale=0.16]{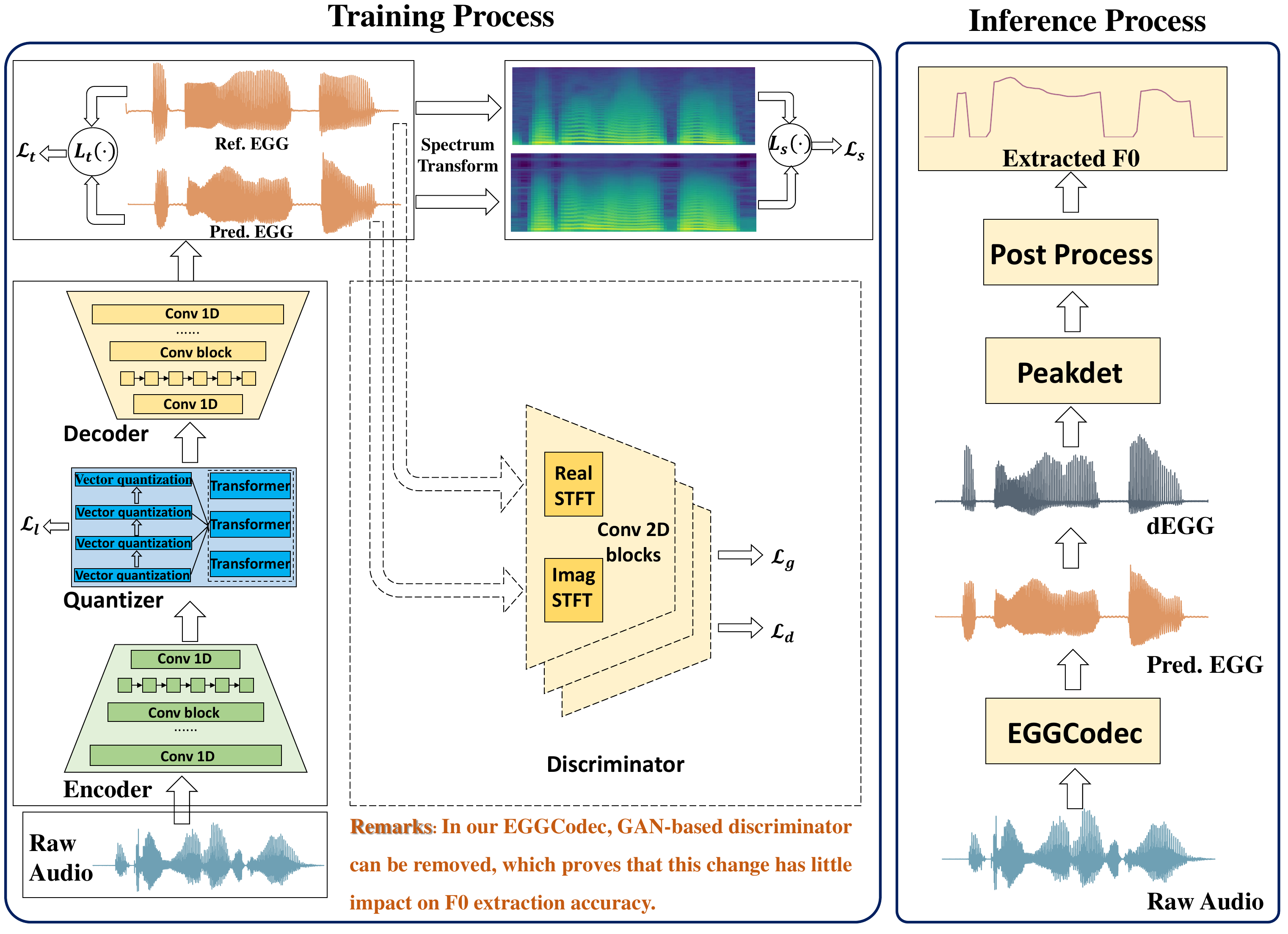}
\vspace{-1.0em}
\caption{The framework of the proposed EGGCodec.}
\label{fig1}
\vspace{-1.5em}
\end{figure}

\section{THE DEVELOPED EGGCODEC FRAMEWORK}

\par In this section, we provide a detailed description of the proposed EGGCodec. From Fig. \ref{fig1}, it shows the input signals during the training process, including the raw audio signal, the lable signal (i.e., Ref. EGG), and our output signal (i.e., Pred. EGG). Next, we introduce EGGCodec in three parts.

\vspace{-1em}

\subsection{Transformation of the Reconstruction Target}

\par The proposed EGGCodec can not only process speech signals but also synchronously collect EGG signals. It introduces an additional 3-7 dB of white noise into the speech signals during training to augment the data. The main estimation target of our model is to reconstruct the EGG waveform from original speech signals for reliable F0 extraction. In EGGCodec, the speech signals are firstly encoded into compact representations, which are subsequently quantized by the quantizer and then reconstructed into waveforms by the decoder. By calculating the error between output signals and target signals, EGGCodec transfers the reconstruction target from speech signals to EGG signals.  The reason why it can be achieved is that the loss function that measures the differences between the predicted and target signals is used to update the model's parameters during the training process. By using the EGG signal as the target of the loss function, the model learns to generate outputs that closely match EGG signals rather than speech signals. This helps our proposed EGGCodec focus on reconstructing the vocal cord vibration signal from the speech input, to more accurately capture F0's fine details. This approach not only improves the accuracy of F0 extraction but also enhances the ability of EGGCodec to characterize the dynamics of vocal fold opening and closing. \footnote{The proposed EGGCodec follows the EnCodec framework that consists mainly of an initial convolution, multiple residual and downsampling modules, and a timing modeling component \cite{defossez2022high}.}

\vspace{-1em}

\subsection{The Design of Loss Functions}

\par To ensure fidelity between reconstructed waveforms and target EGG signals across time-frequency domains, a multi-scale frequency-domain loss function that integrates L1 and L2 norms is proposed to access Mel-spectrogram differences across varied window lengths, complemented by a time-domain correlation loss to improve generalization. This approach enhances EGGCodec's capability to effectively capture multi-scale information features, which are represented as follows:

\vspace{-1.2em}

\begin{multline}\label{eq:freq_loss}
\mathcal{L}_{s} = \frac{1}{6} \sum_{i=5}^{10} \left( \left\| \mathrm{S}(y_{\mathrm{pred}}, 2^i) - \mathrm{S}(y_{\mathrm{ref}}, 2^i) \right\|_1 \right. \\
\left. + \left\| \mathrm{S}(y_{\mathrm{pred}}, 2^i) - \mathrm{S}(y_{\mathrm{ref}}, 2^i) \right\|_2 \right),
\end{multline}
where $y_{\mathrm{pred}}$ and $y_{\mathrm{ref}}$ represent the reconstructed waveform and the target EGG waveform, respectively. \(\mathrm{S}(y, w)\) is the log-Mel spectrogram of signal \(y\) derived from the STFT with window length \(w\). In Eq. (\ref{eq:freq_loss}), we employ a linear combination of L1 and L2 losses applied to the Mel-spectrogram, computed across multiple frequency windows to ensure spectral consistency. In the time domain, EGGCodec further improves the preservation of the signal's phase information by introducing a cosine distance loss, which can be represented by

\vspace{-0.5em}

\begin{equation}\label{eq3}
   \mathcal{L}_{\text{cos}} = 1 - \frac{\langle \mathbf{y}_1, \mathbf{y}_2 \rangle}{\|\mathbf{y}_1\| \|\mathbf{y}_2\|},
\end{equation}
where \( \mathbf{y}_1 \) and \( \mathbf{y}_2 \) are two arbitrary time series. The cosine distance assesses their similarity via the normalized inner product, rendering it scale invariant. A lower value indicates a higher degree of correlation between the two sequences, whereas values closer to 1 indicate weaker similarity. Unlike Euclidean-based metrics that emphasize absolute amplitude disparities, cosine distance prioritizes relative patterns. To further enhance reconstruction accuracy, a hybrid time-domain loss function that combines L1, L2, and cosine distance is formulated as follows:

\vspace{-1em}

\begin{equation}
   \mathcal{L}_{t} = \frac{\mathcal{L}_{\mathrm{L1}}(y_{\mathrm{pred}}, y_{\mathrm{ref}}) + \mathcal{L}_{\mathrm{L2}}(y_{\mathrm{pred}}, y_{\mathrm{ref}})}{\lambda} + \mathcal{L}_{\text{cos}}(y_{\mathrm{pred}}, y_{\mathrm{ref}}).
\end{equation}
where $\lambda$ denotes the weighting factor, empirically
set to 100, to balance the contributions of the different loss terms, ensuring stable training and optimal performance. The terms $\mathcal{L}_{\mathrm{L1}}(y_{\mathrm{pred}}, y_{\mathrm{ref}})$ and $\mathcal{L}_{\mathrm{L2}}(y_{\mathrm{pred}}, y_{\mathrm{ref}})$ quantify absolute and squared differences, respectively, while $\mathcal{L}_{\text{cos}}(y_{\mathrm{pred}}, y_{\mathrm{ref}})$ enhances them with a scale-invariant similarity metric. To balance their contributions, we introduce a scaling factor of 100 to the L1 and L2 losses, empirically derived from extensive trials. This factor establishes a practical normalization magnitude, preventing dominance by L1 and L2 losses while preserving both absolute and pattern-based similarities fidelity.

\par Furthermore, we assign different weights to the aforementioned loss functions and combine them to form the reconstruction loss function, as follows:

\begin{equation}\label{eq4}
   \mathcal{L}_{\text{reco}} = \mathcal{L}_{s} + \lambda \times \mathcal{L}_{t} + \mathcal{L}_{g} + \mathcal{L}_{d} + \mathcal{L}_{l},
\end{equation}
where \(\mathcal{L}_{g}\) embodies the generator's adversarial loss, ensuring perceptual fidelity in the reconstructed signal, while \(\mathcal{L}_{d}\) reflects the discriminator's adversarial loss, distinguishing real from synthesized signals. Additionally, \(\mathcal{L}_{l}\) denotes the entropy coding loss, optionally employed with a Transformer-based language model to enhance compression efficiency \cite{defossez2022high}. The coefficient of $\lambda = 100$, empirically tuned for \(\mathcal{L}_{t}\), balances the loss terms, mitigating magnitude disparities between \(\mathcal{L}_{s}\) and \(\mathcal{L}_{t}\) that could destabilize training. Rigorous optimization confirms this factor fosters both stability and peak performance.

\vspace{-0.8em}

\begin{figure}[h]
\centering
\includegraphics[scale=0.7]{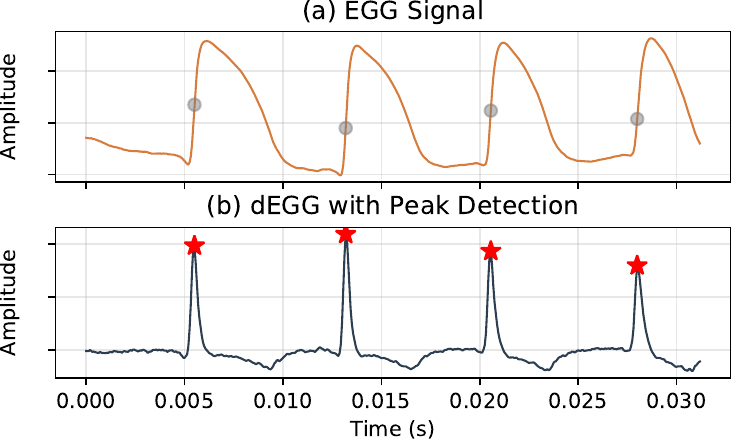}
\vspace{-1.0em}
\caption{The F0 extraction from dEGG signals using the peakdet algorithm to highlight peaks.}
\label{fig_dEGG}
\vspace{-0.5em}
\end{figure}

\begin{figure*}[t]
    \centering
    \begin{minipage}[b]{0.3\textwidth}
        \centering
        \includegraphics[scale=0.36]{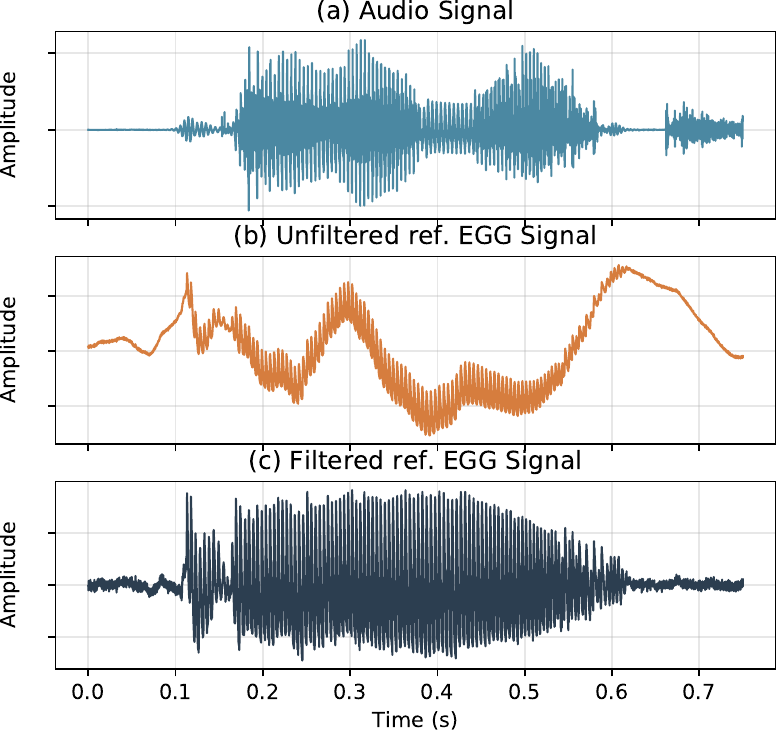}
        \caption{Comparison of unfiltered/filtered EGG signals and the need of processing for the PTDB-TUG dataset.}
        \label{fig:filter}
    \end{minipage}
    \hfill
    \begin{minipage}[b]{0.3\textwidth}
        \centering
        \includegraphics[scale=0.36]{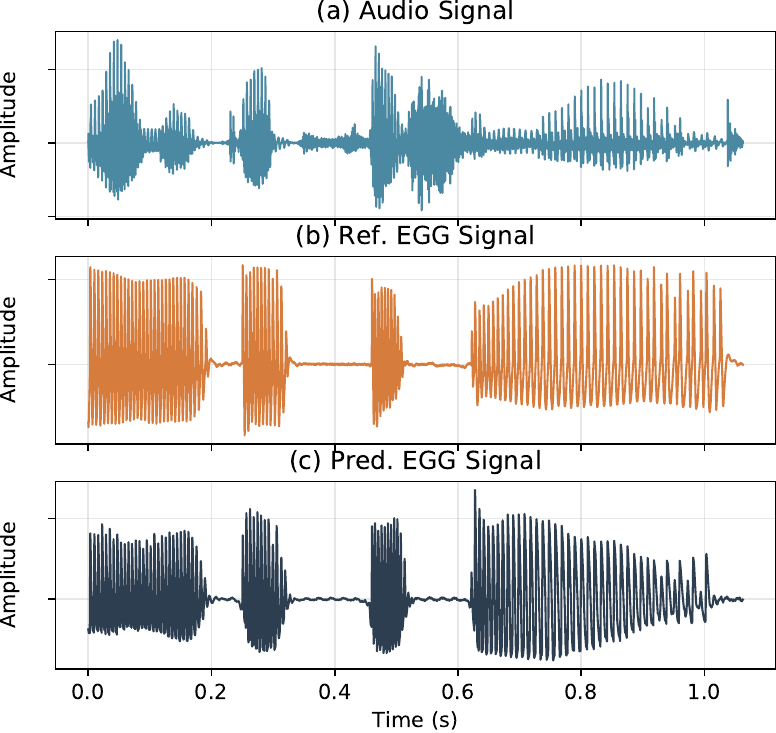}
        \caption{Amplitude comparison: Reconstructed vs. original EGG and audio signals.}
        \label{fig:audio_refEGG_predEGG}
    \end{minipage}
    \hfill
    \begin{minipage}[b]{0.3\textwidth}
        \centering
        \includegraphics[scale=0.36]{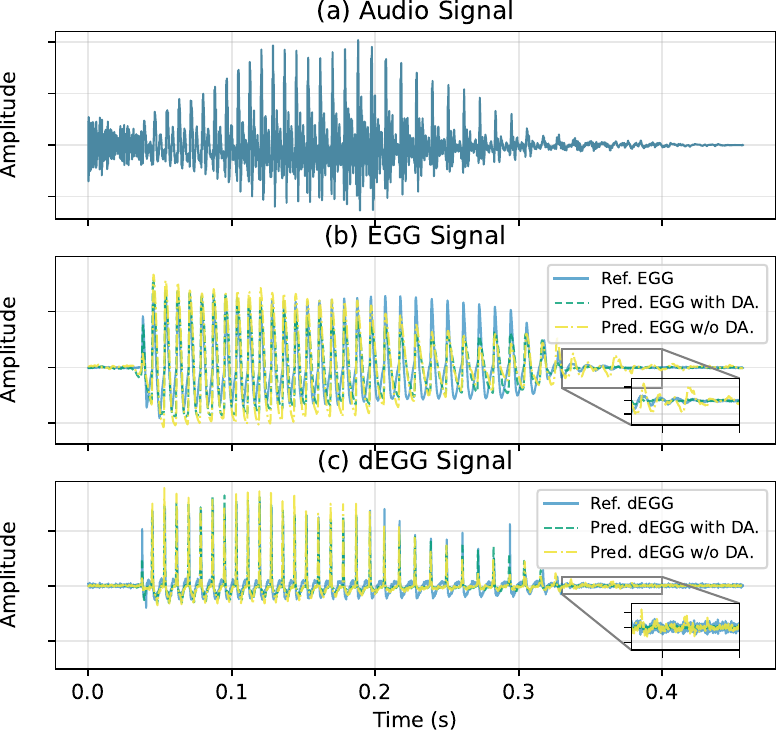}
        \caption{EGGCodec reconstruction: dEGG peaks without noise augmentation.}
        \label{fig:no_noise_aug}
    \end{minipage}
\end{figure*}

\subsection{The Design of F0 Extraction Scheme}

\par In EGGCodec, we propose an encoder-decoder architecture for accurate EGG reconstruction, combined with differential EGG (dEGG) techniques to enhance F0 extraction. The reconstructed EGG signal is differentiated to generate the dEGG signal, as shown in Fig. \ref{fig_dEGG}, where peaks correspond to vocal fold closure instants. Leveraging the peakdet algorithm \cite{henrich2004use,michaud2005final,mazaudon2009tonal,michaud2004measurement}, these peaks are detected as periodic markers to calculate vibration periods and derive F0. Frame-level frequency predictions are subsequently generated through post-processing. By exploiting the alignment between dEGG signal maxima and vocal fold dynamics, EGGCodec ensures robust and accurate F0 estimation. The necessity of processing EGG signals from the PTDB-TUG dataset \cite{pirker2011pitch} is demonstrated in Fig. \ref{fig:filter} \footnote{Notably, the PTDB-TUG dataset is only used in the training phase of EGGCodec. It provides large-scale, synchronously aligned speech and EGG signals suitable for supervised learning and robust training. In the evaluation part, EGGCodec uses the CSTR-FDA dataset that has clinically validated fundamental frequency reference trajectories and is a widely used gold standard corpus in speech research \cite{bagshaw1993enhanced}.}. As revealed in Fig. \ref{fig:filter} (b), unfiltered reference EGG signals exhibit substantial low-frequency components originating from throat muscle artifacts during speech production rather than vocal fold vibrations. These extraneous components, which bear minimal relevance to fundamental frequency extraction, risk interfering with model training. Fig. \ref{fig:filter} (c) further illustrates how omitting the 50 Hz high-pass filter permits these low-frequency artifacts to disrupt the learning process. Our performance evaluation systematically benchmarks filtered versus unfiltered EGG references to validate this critical preprocessing step.

\section{Performance Evaluations}

\subsection{Parameter Settings and Dataset}

\par EGGCodec is trained using the Adam optimizer with the learning rate of $\eta \!=\! 10^{-3}$, the momentum coefficients $\beta_1 \!=\! 0.9$ and $\beta_2 \!=\! 0.999$, and $\epsilon \!=\! 1 \times 10^{-8}$ \cite{kingma2014adam}, and it is trained for 20 epochs using batched processing (14 audio clips per batch), accelerated by an NVIDIA RTX 4090 GPU. The PTDB\_TUG corpus \cite{pirker2011pitch} is exploited as training dataset, which comprises 576 minutes of English speech from 20 speakers (10 male/10 female) with synchronized EGG recordings. To ensure multi-condition robustness, EGGCodec implements additive white noise augmentation across four SNR levels: 3 dB, 5 dB, 7 dB, and clean. EGGCodec's performance is evaluated on the CSTR-FDA dataset \cite{bagshaw1993enhanced}, a gold-standard pitch determination corpus containing 5.53 minutes of speech (1 male/1 female) with ground-truth frequency contours. We calculate the PPMCC to assess the linear correlation between waveforms, which can be defined in reference to \cite{nahler2009pearson,10081490, 10800253}. In Table \ref{tab:performance}, PPMCC measures the linear correlation between the reconstructed EGG signal and the reference EGG signal. A higher PPMCC indicates that the reconstructed EGG signal more closely matches the original, suggesting that the model better preserves the periodic characteristics of vocal fold motion. This is crucial for F0 estimation, as F0 is extracted from the reconstructed EGG signal. If the reconstructed EGG is distorted, the accuracy of the F0 estimation may be compromised. We benchmark EGGCodec against pYIN \cite{mauch2014pyin}, Crepe \cite{kim2018crepe}, dio\_stone \cite{morise2009fast}, and Wav2F0 \cite{10800188}, evaluating mean absolute error (MAE), 50-cent raw pitch accuracy (RPA) \cite{10250914}, 20\% gross pitch error (GPE) \cite{sukhostat2015comparative}, voicing decision error (VDE) \cite{10250914}, and PPMCC.

\vspace{-1em}

\subsection{Experimental Results}

\par As depicted in Fig. \ref{fig:audio_refEGG_predEGG}, we compare the reconstructed EGG signals obtained by EGGCodec with the original EGG signals. The reconstructed EGG signals (Fig. \ref{fig:audio_refEGG_predEGG} (c)) exhibit a high degree of consistency with the original signals (Fig. \ref{fig:audio_refEGG_predEGG} (b)) in the vibrating regions of the vocal cords. In the voiced regions of the audio signal, the reconstructed EGG signal successfully avoids generating erroneous waveforms when the original EGG signal shows no significant vibration, suggesting the reconstruction process effectively reduces the interference from non-vocal vibrations. Moreover, the reconstructed EGG waveforms maintain nearly the same clarity and simplicity as the original signals, preserving their essential characteristics while reducing potential distortions. This faithful reconstruction preserves essential details for F0 extraction while reducing noise and irrelevant information, providing a cleaner, more reliable foundation for F0 extraction. As shown in Fig. \ref{fig:no_noise_aug}, the effect with or without (w/o) noise-augmented training on EGG reconstruction is investigated. From Fig. \ref{fig:no_noise_aug} (b) and (c), it can be seen that training without noise augmentation increases the reconstructed signal’s vulnerability to noise during unvoiced segments, increasing the risk of misclassifying noise as F0. In contrast, noise-enhanced training yields perfect reconstructed EGG signals, enabling accurate vocal fold cycle detection. Experimental results reveal that the stability of the reconstructed signal is essential for accurate peak detection. Incorrect EGG reconstructions yield abnormal dEGG peaks, which potentially cause segmentation errors in vibration cycles and ultimately affect the accuracy of F0 extraction.

\vspace{-0.5em}

\begin{table}[h]
\centering
\footnotesize
\setlength{\tabcolsep}{2.0pt}
\caption{Performance comparison of the EGGCodec with the state-of-the-art baselines and its ablation studies.}
\label{tab:performance}
\begin{tabular}{l S[table-format=1.3] S[table-format=3.2] S[table-format=2.1] S[table-format=2.1] S[table-format=2.1]}
\toprule
\multirow{2}{*}{Model Name} & {PPMCC} & {MAE} & {RPA} & {GPE} & {VDE} \\
& {$\uparrow$} & {(Hz) $\downarrow$} & {(\%) $\uparrow$} & {(\%) $\downarrow$} & {(\%) $\downarrow$} \\
\midrule
dio\_stone \cite{morise2009fast} & {-} & \textbf{14.14} & \textbf{88.1} & 8.3 & \textbf{8.9} \\
pYIN \cite{mauch2014pyin} & {-} & 36.85 & 62.9 & 24.3 & 26.1 \\
crepe \cite{kim2018crepe} & {-} & 16.10 & 87.8 & \textbf{8.0} & 9.5 \\
 Wav2F0 \cite{10800188} & {-} & 15.18 & 81.8 & 9.7 & 8.2 \\
(EGGCodec, Optimal) & 0.834 & \textbf{13.69} & 86.0 & 9.1 & \textbf{5.5} \\
(EGGCodec, Cos) & 0.818 & \textbf{15.76} & 85.4 & 10.1 & \textbf{5.9} \\
(EGGCodec, L1/L2) & 0.468 & 54.74 & 72.9 & 24.8 & 23.2 \\
(EGGCodec, w/o Time)  & 0.002   & 40.30  & 56.6  & 35.8   & 9.2 \\
(EGGCodec, w/o Freq)   & 0.82   & \textbf{15.31}  & 86.5  & 10.2   & \textbf{6.0} \\
(EGGCodec, 5dB NDA) & 0.828 & 16.68 & 84.7 & 10.9 & \textbf{6.5} \\
(EGGCodec, w/o NDA) & 0.819 & 17.27 & 84.1 & 12.2 & \textbf{7.4} \\
(EGGCodec, 7dB NDA) & 0.839 & 17.41 & 84.7 & 11.1 & \textbf{6.5} \\
(EGGCodec, w/o GAN) & 0.812 & 14.17 & 86.1 & 9.5 & \textbf{5.5} \\
(EGGCodec, Unfiltered) & 0.278 & 26.71 & 73.9 & 19.0 & 10.3 \\
\bottomrule
\end{tabular}
\end{table}

\par To systematically evaluate each EGGCodec component's contribution, we designed multiple control groups for ablation experiments. The \textbf{(EGGCodec, Optimal)} configuration integrates cosine distance and L1/L2 norm losses in the time domain for robust error characterization, employs L1/L2 losses in the frequency domain to capture spectral features, and uses 3 dB SNR white noise augmentation with GAN-based adversarial training to enhance noise resilience and fidelity. A 50 Hz high-pass filter is applied to the reference EGG signal to remove low-frequency interference while preserving critical features, achieving a better balance between accuracy and robustness across evaluation metrics. Other control groups include: \textbf{(EGGCodec, Cos)} and \textbf{(EGGCodec, L1/L2)} solely consider cosine distance and L1/L2 losses in time domain, respectively; \textbf{(EGGCodec, w/o Time)}, sans time-domain losses; \textbf{(EGGCodec, w/o Freq)}, sans frequency-domain losses; \textbf{(EGGCodec, w/o NDA)}, \textbf{(EGGCodec, 5dB NDA)}, and \textbf{(EGGCodec, 7dB NDA)}, varying noise augmentation (none, 5 dB SNR, 7 dB SNR); \textbf{(EGGCodec, w/o GAN)}, excluding GAN training; and \textbf{(EGGCodec, Unfiltered)}, assessing the effectiveness of high-pass filtering.

\par Table \ref{tab:performance} presents a comparative performance analysis of EGGCodec against baselines. \textbf{(EGGCodec, Optimal)} achieves outstanding performance in F0 extraction, with an MAE of just 13.69 Hz, outperforming all baselines. \textbf{(EGGCodec, Optimal)} significantly outperforms pYIN, which records a much higher MAE of 36.85 Hz, achieving the lowest VDE at 5.5\%, indicating superior robustness in voiced/unvoiced classification. Among control groups, \textbf{(EGGCodec, w/o GAN)} achieves an MAE of 14.17 Hz and an RPA of 86.1\%. \textbf{(EGGCodec, Cos)} and \textbf{(EGGCodec, L1/L2)} demonstrate competitive performance, with the former maintaining a relatively low MAE of 15.76 Hz. \textbf{(EGGCodec, Optimal)} and \textbf{(EGGCodec, 7dB NDA)} achieve the highest PPMCC values, i.e., 0.834 and 0.839, respectively, indicating superior EGG reconstruction and providing a more reliable foundation for F0 extraction. In contrast, \textbf{(EGGCodec, Unfiltered)} has a significantly lower PPMCC of 0.278, suggesting a poor correlation with the reference EGG, which may negatively impact subsequent F0 extraction. Thus, EGGCodec not only enhances the accuracy of EGG reconstruction but also contributes to the stability and reliability of F0 extraction. It can also be observed that \textbf{(EEGCodec, 7dB NDA)} presented PMCC slightly higher than \textbf{(EEGCodec, Optimal)}. This is mainly because PMCC measures the linear correlation of the waveform as a whole and is not affected by absolute amplitude offset and scale variation. Moreover, MAE and GPE are based on a point-by-point comparison of the F0 values of the reconstructed signal with the reference signal. The difference between the metrics is more likely to reflect the different responses of the F0 extraction to the microstructure of the signal rather than a degradation of the overall reconstruction quality.

\vspace{-1.5em}

\section{CONCLUSION}
This letter proposes an innovative neural Encodec framework named EGGCodec, which advances EGG signal reconstruction and F0 extraction. EGGCodec delivers exceptional accuracy and generalization on an EGG-rich dataset. Comparisons with state-of-the-art F0 extraction schemes highlight EGGCodec's significant gains, and ablation studies demonstrate the contribution of each component to its performance. In our future work, we intend to explore the F0 estimation results of the proposed EGGCodec under urban acoustic noise and different SNRs, in order to further demonstrate the robustness of EGGCodec to noise.

\footnotesize
\bibliographystyle{IEEEtran}
\bibliography{IEEEabrv,ref}
\end{document}